# Complex Cognition: A New Theoretical Foundation for the Design and Evaluation of Visual Analytics Systems[1]

Xiaolong (Luke) Zhang (xuz14@psu.edu)
College of Information Sciences and Technology, Penn State University, University Park, 16802

**Abstract:** Current research on visual analytics systems largely follows the research paradigm of interactive system design in the field of Human-Computer Interaction (HCI), and includes key methodologies including design requirement development based on user needs, interactive system design, and system evaluation. However, most studies under this paradigm have a contradiction: there is a significant mismatch between the research methods developed for simple cognitive behaviors (e.g., color perception, the perception of spatial relationship among interactive artifacts) and research goals targeting for complex analytical behaviors (e.g., reasoning, problem-solving, decision-making). This mismatch may hurt the theoretical contributions of research studies, in particular the internal validity of a designed system and the external validity of design methods. To address this challenge, this paper argues for a need to go beyond traditional HCI theoretical foundations and proposes to adopt complex cognition theories to build new theoretical foundations. Specifically, this paper analyzes how current design and evaluation methods in research on visual analytics systems constrain the internal and external validity of research, discusses the connections between complex cognition theories and visual analytics tasks, and explores how problem-solving theories from complex cognition can guide research on visual analytics systems.

---

[1] Translated by the author from the original Chinese paper presented at ChinaVIS 2025.

## 1. Introduction

Currently, in the field of visual analytics (VA) research, system design is a mainstream direction. In major international conferences such as IEEE VIS, EuroVis, and PacificVis, this type of research accounts for a significant proportion. These works generally follow the research paradigm of interactive system design in the field of Human-Computer Interaction (HCI), which includes key stages such as user requirement analysis, system design, and system evaluation. First, design requirements are formulated based on the challenges faced by users or potential users in specific data analysis domains; then, a visual analytics system is implemented to meet these requirements; finally, user evaluation is conducted to verify whether the system has achieved its intended goals in terms of both system functionality and user experience.

However, a contradiction exists in most studies under this paradigm: there is a significant mismatch between research methods and the actual behavior of target users. On one hand, the methods relied upon for requirement analysis and evaluation in the HCI field are typically aimed at simple cognitive behaviors (e.g., color perception, identification of spatial relationships between interactive components) and deterministic interaction tasks (e.g., selecting specific data from an interface). In contrast, behaviors supported by VA systems—such as reasoning, problem-solving, and decision-making—transcend the scope of simple cognition and are characterized by task indeterminacy and the need for dynamic strategy adjustment. On the other hand, the metrics used in system evaluation are also largely derived from the HCI field, often targeting usability indicators like efficiency and accuracy for individual tools and user experience, while neglecting the effectiveness of support for complex analytical behaviors. This paradigm mismatch between methods based on simple cognitive behaviors and research goals oriented toward complex cognition weakens the theoretical depth of the research, affecting, for example, the internal validity (effectiveness of the system) and the external validity (generalizability of design methods).

Solving this mismatch requires breaking through traditional HCI theories and methods and constructing a new theoretical foundation for VA system design by integrating scientific theories related to analytical behavior. This paper proposes Complex Cognition Theory as the new theoretical foundation to resolve the limitations of existing methods. Specifically, this work includes: first, summarizing current design and evaluation methods in VA and analyzing their constraints on internal and external validity; second, elucidating the correlation between complex cognition theory and VA tasks; and finally, exploring how to utilize problem-solving theories within the domain of complex cognition to guide related system research.

## 2. Methods and Means of Visual Analytics System Design

The research process for VA systems generally includes several major steps: user requirement analysis, system design, and system evaluation. The goal of requirement analysis is to summarize user needs and propose corresponding design objectives based on an understanding of the characteristics and "pain points" of user work. System design is the specific implementation of these design goals, often involving data processing algorithm tools, visualization tools, and interaction tools. System evaluation involves testing the designed system prototype to examine whether the system meets the design requirements.

## 2.1 Requirement Analysis

Requirement analysis provides the objectives and basis for system design. Methods for requirement analysis can basically be divided into two categories: those based on literature and those based on actual empirical research. The former can be applied to some common and universal analytical problems (such as social network data analysis or spatio-temporal data analysis). For such problems, literature is usually abundant, and user requirements can basically be determined through systematic literature review. Requirement analysis based on actual investigation often targets user behaviors that are relatively new and lack sufficient prior research. This section focuses on investigation-based requirement analysis: questionnaires, user interviews, and field observation.

### 2.1.1 Questionnaires

The questionnaire method obtains behavioral data from target users by distributing a unified questionnaire to a participant group. A questionnaire usually includes multiple questions, and the statistical analysis of questionnaire data can reveal the general behavioral characteristics of a user group, such as representative behaviors when performing certain types of tasks or representative opinions on certain tools. In a study of a visual analytics system using machine learning to support the labeling of event information in sports videos [1], the user requirement research utilized questionnaires to understand the events of interest to fans when watching tennis, table tennis, and badminton matches. The questionnaire was distributed to approximately 600 members of two different online communities, and the results helped determine the video information features that the machine learning algorithm needed to focus on.

The advantage of the questionnaire method is the superior external validity of its results. If the participant group of a questionnaire has a sufficient sample size and broad representativeness, its results can be applied to other similar situations. Therefore, a questionnaire should be distributed to as many different communities and involve as many user groups as possible.

### 2.1.2 User Interviews

User interviews collect required information through dialogue with users. The focus of interviews is to allow users to provide the information needed for research from their own perspective, such as how they perform a task or use a system. Interview subjects are often multiple representative users, and the interview process can be fully structured or semi-structured. In structured interviews, the questions asked and their order are set in advance and cannot be changed; in semi-structured interviews, only some key questions are pre-determined, and researchers can control the order of questions and pose new questions on the spot based on the participant's answers.

For example, in the aforementioned research on the sports video labeling system, requirement analysis also employed the interview method to understand the information of interest and the difficulties encountered by expert users (such as sports researchers and sports data analysts) when analyzing match videos.

The advantage of the interview method is the ability to obtain accurate, in-depth, and subjective information directly from users through communication. However, the success of interview results depends on several factors, such as the design of interview questions, the participation of representative users, and the control of the interview process.

### 2.1.3 Field Observation

Field observation is often conducted at the user's actual workplace with the goal of observing the characteristics displayed and problems encountered by users when performing actual tasks. While questionnaires and interviews primarily aim to obtain internal behavioral information related to the user, the focus of field observation is on external environmental factors that may influence user behavior (such as workspace conditions and interactions with colleagues).

The users and tasks targeted by field observation also need to be representative, and researchers can have different degrees of intervention during the observation process: as a pure observer without any intervention; as an observer who asks questions about certain user behaviors; or even by directly participating in the execution of the task. Different approaches have their pros and cons. As an observer, the researcher can avoid affecting the user's actual operations and work efficiency; but as a participant, the researcher can gain a deeper understanding of the user's experience and needs during operation through personal experience.

For example, in a study regarding a visual analytics system for autonomous vehicle driving systems [2], requirement analysis utilized field observation of expert users' work to understand the modules they care about, the parameters they rely on, and the "pain points" they encounter when evaluating autonomous driving systems, which helped establish the design requirements and goals of the system.

The advantage of field observation lies in its ability to capture user behavior data in real scenarios. When users perform routine tasks in a natural work environment, their behavior often possesses high reliability. However, interference from non-target behaviors may occur during field observation, thereby introducing data noise and increasing the difficulty of data analysis.

## 2.2 System Design

System design generally takes the results provided by requirement analysis as the design goal, involving technical development and implementation in terms of data processing tools, visualization tool design, and interaction tools.

### 2.2.1 Data Processing Tool Design

The primary purpose of data processing tool design is to develop algorithms or methods for a specific design goal. Raw data faced by visual analytics systems often has characteristics such as large volume, loose structure or missing values, and mismatch between inherent data structures and task-required structures, making it difficult to visualize directly according to task requirements. Therefore, it is necessary to rely on data processing algorithms or methods to pre-process the raw data (such as extracting useful data and structuring data) to facilitate subsequent visualization design.

In many cases, the required algorithms already exist, and the focus of design work is on implementing the algorithm for the existing data. For example, in a study of a visual analytics system supporting online communication patterns of MOOC students [3], the primary analytical task required extracting relevant communication topics from students' text exchanges. Since topic extraction algorithms are already mature in the field of data mining, the work required here was to select and implement a suitable algorithm and integrate it into the system.

However, in some cases, it is also necessary to develop new algorithms or methods based on task requirements. This type of work possesses a certain degree of innovation in the field of visual analytics; therefore, many VA studies provide detailed descriptions of relevant new algorithms. For instance, in a visual analytics study targeting concept drift in data [4], the work on data processing tools included the design and implementation of a new method for detecting concept drift.

### 2.2.2 Visualization Design

Visualization design here refers to the design of various views within the visual analytics system. A VA system often contains multiple views, and the design of each view is essentially consistent with traditional information visualization design, which targets a given data structure type and specific tasks to design a particular view. In the aforementioned MOOC communication data VA system, there is a network view showing the characteristics of student communication relationships to help users observe these relationships at various levels. This network visualization tool was implemented based on the traditional force-directed graph method.

However, because a VA system has multiple views, research literature can only provide simple descriptions of the design process for such views and cannot provide detailed descriptions of the design process—such as comparisons and selections of different design schemes—as found in information visualization design research literature; these details are generally ignored. For example, in a VA system for social networks [5], the system features a novel "TreeNetViz" component used to support the analysis of a new data structure that merges tree and network structures. But since this view is only one part of the overall analysis system, the article's introduction of it focuses on its functional role within the system, while the specific design work for the view (such as design philosophy and related algorithms) only cites the corresponding visualization design literature [6].

Some research literature does indeed include novel view designs [7], but because the focus is on visual analytics rather than visualization design, the effectiveness of the design is not evaluated.

### 2.2.3 Interaction Design

System interaction design includes interaction between users and certain visual design elements in a view, task-based interaction across multiple views, and interaction design that supports the analytical workflow. Interaction with certain visual design elements is often limited to a single view and in most cases falls within the scope of basic interaction tasks [8]. Implementation of interactions supporting these universal tasks can utilize the inherent functions and interfaces of development toolkits. For example, many internal tools in d3.js can easily implement functions like data filtering and highlighting.

Interaction design for specific analytical behaviors requires a deep understanding of the analysis task itself to optimize the overall interface layout and synchronization between views. For example, if the algorithms and views commonly used by target users when performing a task are understood, the overall interaction design of the interface can display the involved algorithms and views simultaneously rather than letting users configure them themselves. At the same time, designing these related views as multiple coordinated views allows users to quickly and effectively obtain information relevant to the task analysis. Similarly, coordination between views is necessary to support complex analysis tasks. For example, VA systems often require coordination among multiple views so that when a user selects certain data points in one view, these data points should be distinguished from other data in other views, allowing users to conveniently observe the characteristics of the same set of data from multiple perspectives. Such designs need to consider the characteristics of the task itself.

Interaction design also considers support for the analytical workflow, with the goal of comprehensively supporting various tasks in the user's analysis process and the transitions between them. Currently, much of this work relies on analysis models such as sensemaking. These models propose major stage tasks in visual analytics and the relationships between them (such as presentation and iteration between tasks). According to these models, the design of VA systems can consider corresponding tools to support each task and the transitions between tasks. For example, Pirolli and Card's sensemaking model [9] includes multiple interlinked modules; for the three modules of "shoebox," "evidence file," and "schema," corresponding data collection tools can be provided in the design to facilitate users in collecting relevant data for the three modules independently. At the same time, the model also points out the links between the three modules—such as quickly browsing data in the shoebox module to provide evidence for the evidence module, or assisting users in obtaining important relationship patterns by structuring data in the evidence module—and interaction design can consider corresponding tools to support such transformation and progression between analysis steps.

## 2.3 System Evaluation

A visual analytics system is often verified through evaluation to see if the design meets the design goals. Currently common evaluation methods include case studies, user interviews, and experimental evaluation. We analyzed articles on VA system research published in the three major visualization conferences (IEEE VIS, EuroVis, PacificVis) from 2007 to 2021. Among a total of 230 articles, we found that about 48% used case studies in their evaluation work, about 32% used user interviews, but less than 7% employed experimental evaluation methods. Data for articles after 2021 has not been fully analyzed, but preliminary results show that articles in recent years exhibit a pattern similar to those between 2007 and 2021.

### 2.3.1 Case Studies

Case study is a common research method in behavioral science [10], focusing on exploring a new phenomenon or describing a complex phenomenon through a typical case. However, the case study method cannot verify causal relationships, such as the impact of a tool or design on user behavior.

Case study evaluation methods in VA system research generally demonstrate system utility by describing the actual situation of users performing analytical work with the designed system. They often focus on describing which tools users used, what steps they took, and what problems they encountered when using a VA system. Usually, descriptions of case studies rely on indicators such as task completion and user satisfaction to prove the system's effectiveness. For example, in a study of a map-based VA system for Weibo information [4], the evaluation work used two case studies (one social event, one political event), detailing how the system supported the analysis of the evolution and development of these events through Weibo information, demonstrating how the system helped users quickly identify key figures, locations, and times, and supported the system's usability and effectiveness through user feedback.

### 2.3.2 Interviews

Interviews obtain users' subjective evaluations of the designed system through direct communication. In VA research, interviews are often combined with case studies. After users have used and experienced the designed system, interviews are conducted to gain an in-depth understanding of their experiences and feedback, providing detailed evidence of the system's impact and role in their work. For example, in the previously mentioned VA system research for autonomous vehicle driving systems [7], the evaluation work included an interview session with expert users. This session was conducted after users had used the designed system tools, and user evaluations touched upon the correctness and intuitiveness of the tools and analysis processes provided by the system, as well as the comprehensibility of each view in the system.

### 2.3.3 Experimental Evaluation

In interaction design, the experimental evaluation method essentially assesses the effectiveness of a design. In the fields of visualization and visual analytics, this method compares the merits of tools or systems by comparing the performance of users when using different tools or systems to perform the same task. Since the goal of the experimental evaluation method is to verify the causal relationship between the tools or systems used and the improvement of task performance, the design and implementation of experiments require strict control over other factors that might interfere with this causal effect, such as differences between groups of participants and consistency in performing tasks under different conditions. However, this strict control over other factors also limits the scope of application for experimental methods.

For the evaluation of VA systems, the greatest challenge for the experimental method lies in the complexity of experimental tasks. If the object of evaluation is only a view for a simple task, experimental design is relatively easy. For example, when using experimental methods to test the effectiveness of an adjacency matrix view in helping users understand the network characteristics of a community, the experimental design can contrast user performance when using an adjacency matrix design versus a traditional node-link design. Experimental tasks can be typical network analysis tasks such as finding sub-communities within a community or determining the relationship between two nodes in a community. Such experimental tasks are clear, the level of interaction between users and views is roughly the same, and the design and process of the experiment are relatively easy.

But when the object of evaluation is a VA system for a complex task, experimental design becomes very difficult. For example, when evaluating a VA system oriented toward artificial intelligence models, typical tasks may require users to compare the characteristics of several models from different perspectives, and the goals of such tasks may be vague (which features to compare, how many models to compare, etc.), and the task paths chosen by users may also differ (some users analyze by model unit, while others base it on data features). If these differences are not effectively controlled, the validity of experimental results may be questioned. Strictly controlling these factors would greatly increase the complexity of the experiment and might even lead to the experiment being unable to proceed. If, in order to reduce task path differences, users are forced to follow a specific analysis method, it might force some users to use unfamiliar methods, rendering them unable to complete the task. It is precisely because of the difficulty of experimental evaluation methods that the proportion of VA system research using experimental evaluation methods is low.

When evaluating a VA system, a feasible compromise is to evaluate the core tools of the system rather than testing the overall functionality. The advantage of this method is that it significantly reduces experimental complexity. Taking the previously mentioned sports event VA system as an example: given the difficulty in implementing experimental evaluation for the complete analysis task, the experimental evaluation work selected two core sub-tasks (basic information extraction analysis tools and high-level semantic information integration analysis tools) and designed verification experiments for them respectively.

## 3. Mismatch Between VA Design Methods and Research Objectives

The mainstream methods currently adopted by visual analytics face some challenges, the fundamental reason being a mismatch between the research paradigm of the HCI field and the core goals of VA systems. This section first explains the essential differences between analytical behavior and general interactive behavior by contrasting their characteristics; then, it delves into the main stages of system design to analyze the limitations existing in current VA system research methods.

### 3.1 Simple Interaction Behavior vs. Complex Analytical Behavior

Methodologies in the field of HCI are largely built upon theories of simple cognitive behavior. Such simple cognitive behaviors typically involve rapid response mechanisms like perception, attention, and intuition, with typical manifestations including basic operations like color/shape perception, menu/icon selection, and shortcut key usage. A basic assumption in this field is that users already clearly know the system command to be executed, and executing that command requires some tool or medium provided by the computer system. This assumption is fully reflected in Norman's interaction gap theory: the theory proposes the "gulf of execution," specifically referring to the barrier where a user cannot complete a target task due to the lack of interactive tools, and one goal of interaction design is precisely to eliminate such barriers.

However, the behavior supported by VA systems is not essentially this type of simple behavior; it usually requires high-level cognitive activities such as reasoning, decision-making, learning, and problem-solving. These high-level cognitive activities possess significant non-deterministic characteristics: users need to constantly define tasks and adjust strategies. It is worth noting that tools supporting complex analytical behavior are not a simple overlay of basic interaction functions. Taking Tableau and ArcGIS as examples, although these commercial analysis tools have very user-friendly interface designs at the basic interaction level (such as changing the rendering color of a view element, dragging a visual component, or selecting and highlighting part of the data), being proficient in these tools does not guarantee successful completion of a VA task. In past IEEE VAST challenges, many participating teams used these commercial analysis tools, but the quality of their analytical results varied significantly.

Similar phenomena exist in other interactive systems involving high-level cognitive behavior. For example, the popular Photoshop has very good interaction design and integrates many tools for image processing. But even if proficient in the tool, users may still find it difficult to complete professional design tasks such as high-precision hair masking or movie-level rendering. At its core, the interaction design of these tools generally focuses on universal, general, and simple cognition-related interaction tasks, failing to fully consider the requirements of tasks involving high-level cognitive behavior such as analysis and design.

Visual analytics behavior differs from general simple interactive behavior. VA behavior often has characteristics such as vague goals, uncertain analysis strategies and paths, and task iteration. Personnel performing analysis tasks need to constantly introspect their behavior to examine their analysis process, select corresponding strategies, and constantly adjust strategies based on intermediate results. Obviously, the cognitive model of this behavior is different from the cognitive model of typical interactive behavior in HCI (such as quickly finding an icon or menu item).

The time scale of human action theory proposed by Newell [11] divides human activities into different bands (as shown in Table 1), ranging from millisecond-level biological neural activities (biological band) to social changes spanning years or even longer (social band). Among them, HCI design focuses on the behavioral characteristics of the cognitive band (time scale of 0.1 – 10 seconds) [12], including basic perception-action loops and immediate decision-making processes. In contrast, high-level cognitive activities such as reasoning, decision-making, problem-solving, and learning supported by VA systems belong to the rational band, which has longer time scales (measured in minutes, extending to hours).

Table 1: Time Scales of Human Action (derived from Newell [11])

| Time Scale (seconds) | Band |
|:---:|:---:|
| $10^5$ -- $10^7$ | Social band |
| $10^2$ -- $10^4$ | Rational band |
| $10^{-1}$ -- $10^1$ | Cognitive band |
| $10^{-4}$ -- $10^{-2}$ | Biological band |

Newell and Card [13] pointed out that when the cognitive mechanisms invoked by these two types of behavior are different, their theoretical foundations also differ significantly: traditional HCI behavior primarily follows the basic principles of cognitive psychology, while complex analytical behavior belongs to the category of bounded rationality.

Generally, the core goal of HCI design is to minimize the cognitive resources required by users when completing interaction tasks. For example, when users interact with a system through a command line, they often need to rely on long-term memory to construct command formats and syntax; whereas graphical user interfaces provide interaction elements such as menus and icons, allowing users to directly identify the relevant menu or icon to complete the command. This identification behavior requires far fewer cognitive resources than invoking long-term memory.

Some interaction designs also strive to transform tasks with long time constants into tasks with short time constants. For example, in graphical user interfaces, some animation-based designs (such as dynamically expanding a folder) function at the cognitive level by utilizing object constancy in animation to help users intuitively understand the trajectory of changes in relevant objects (e.g., an object suddenly appearing or changing position) [12], without needing to deploy extra cognitive resources to explain such changes, especially when changes are not actively driven by the user.

From a cognitive theory perspective, the goal of HCI design is to fully utilize the characteristic of human thinking that is good at using experience and intuition to design simple and easy-to-use interactive tools. Cognitive psychology research shows [14] that people use two different cognitive systems in the thinking process: System 1 and System 2. System 1 operates fast and possesses a certain degree of autonomy, its decision-making process primarily based on past experience. What we usually call intuition or subconscious behavior is precisely the action of System 1. However, because System 1's judgment relies on experience and intuition, its decisions may be erroneous, especially when the encountered problem is not completely identical to the past, only being similar on the surface but different in essence.

In contrast, System 2 requires conscious intervention by the brain and deployment of extra cognitive resources such as long-term memory to perform complex tasks like reasoning, problem-solving, and decision-making. Therefore, tasks relying on System 2 usually have higher accuracy rates but slower processing speeds.

In HCI design, the goal is to allow users to primarily rely on System 1 for interaction as much as possible to reduce cognitive load, so users can devote more cognitive resources to more "important" tasks. Analytical behavior usually involves these more "important" tasks, as analysis tasks often cannot be completed by intuition alone and require deployment of System 2 for deep reasoning, problem-solving, and decision-making.

Given the essential differences between typical HCI tasks and VA tasks, we should re-examine current VA research methods and gain a deep understanding of the limitations these HCI-derived methods may have in guiding VA system design, ensuring that designs better meet the needs of the cognitive processes of analytical behavior.

### 3.2 Current Limitations in Requirement Analysis

There is currently a key problem in VA system requirement analysis: a lack of deep understanding and parsing of complex analytical tasks. Based on the aforementioned 230 VA system research papers, we extracted 1,048 requirement analysis items and categorized them according to the targeted interaction behavior (simple interaction behavior or complex analytical behavior). Data shows that about 56% of requirement items focus on simple interaction behavior (such as data filtering, view control, user preference settings, etc.).

However, merely satisfying these basic interaction requirements is not enough to ensure effective support for analysis tasks by the system. These requirements are only the operational foundation in the analysis process, but the core of analysis efficacy lies in higher-level cognitive activities, such as dynamic selection and backtracking of analysis strategies and control of the reasoning process. Therefore, VA system requirement analysis should place its focus on the level of complex task requirements of analytical behavior, rather than the level of basic interaction tasks related to interactive elements like data and views.

### 3.3 Current Limitations in System Design

The main challenge in system design lies in how to define the innovation of this work. VA system research needs innovation in system design, and system design usually includes design work in data algorithms, views, interaction, and system integration; thus, the innovation of system design is often reflected through specific work in these aspects. VA system research literature generally clearly elucidates research contribution points, from which we can extract innovation points of a research project in system design.

Through analysis of 230 relevant papers, we found that when elucidating research contributions, literature rarely lists work in view design, interaction design, and system integration as innovation points. The reason may be that researchers are aware that innovative work in view design belongs to the scope of information visualization, and innovative work in interaction design belongs to the scope of HCI, so both are unsuitable for discussion and evaluation within the scope of VA. Work in system integration mostly belongs to engineering problems and does not belong to the scope of research.

However, many papers take algorithm design as an innovation point. Further analysis indicates that work related to algorithms includes the following two major categories:

- Supporting the target analysis task based on an existing algorithm. This type of algorithm can be directly ported without modification, such as topic models used in systems supporting text information analysis [3], or appropriately modified on the basis of an existing algorithm to meet analysis task requirements;
- Creating new algorithms.

Although creating new algorithms has the highest innovation, the innovation of the algorithm itself transcends the scope that the VA field can judge. For work that implements an existing algorithm, its innovation needs in-depth discussion. Implementing an algorithm identically obviously lacks innovation; for work that improves an existing algorithm, if the improvement only applies to the targeted analysis task and data and is difficult to apply to other tasks and data, then its innovation is also limited. But if the algorithm improvement has strong generalizability, then such work is similar to creating a new algorithm, and its research contribution should be discussed and judged by algorithm-related fields.

View design and interaction design face similar challenges. For view design, if the work only implements existing views, then this aspect lacks innovation. If the proposed design is a brand-new design for a certain task with meaningful generalizability, then its contribution is more suitable for discussion within the field of information visualization rather than VA. For example, the previously mentioned VA system design for social networks [5] used a TreeNetViz component as its core view component. The use of this tree-map visualization in the VA system occurred only after its effectiveness had been verified in the information visualization research field [6] and then integrated into the system.

For interaction design, if the work is limited to supporting general user interaction processes (such as coordination between views, view layout, balance between global and local information, etc.), its innovation is also limited because related tasks have already been extensively studied. But if interaction design work involves new findings representative of a class of analytical behavior (such as interaction between users and algorithms during AI algorithm analysis), then judging the innovation of this work requires not only accurate mastery of the analysis process but also deep understanding of relevant cognitive behaviors and data algorithms; its significance often transcends the VA field and also affects fields like HCI and AI.

So for system design work, how to accurately elucidate the contribution of algorithm design, view design, and interaction design to the VA field is a challenge. If they are discussed separately, a problem we face is whether the VA field has sufficient professional capacity to accurately judge their innovation. As the importance of these three to VA is self-evident, innovation in this work should consider the relationships among the three, which requires a guiding framework that can fuse them together.

### 3.4 Current Limitations in System Evaluation

Research oriented toward system design needs strict evaluation to prove that the designed system can effectively help users complete analysis tasks. However, our analysis of 230 papers shows that this aspect is often the weakest link in research because many evaluation works, strictly speaking, only corroborate system usability through descriptions of system efficacy and feedback from user interviews, rather than adopting credible evaluation means to prove the system's effect on the success of user analysis work. The reason is the selection of inappropriate methods for evaluation, and this limitation in method choice only weakens the contribution of research.

The previous text mentioned that our analysis of literature shows current evaluation means primarily adopt case studies and interviews, with less frequent use of user experiments, despite the latter being the only reliable method to verify the causal relationship between a system and the improvement of analysis behavior effects. Behind this phenomenon lie deep methodological challenges: in HCI and information visualization, although user experiments are widely used to evaluate the improvement effects of specific designs on discrete interaction tasks, traditional experimental methods face multiple dilemmas when facing the multi-component, multi-view, and multi-modal interaction characteristics of VA systems:

- Task complexity: Analysis tasks usually require combining multiple visualization tools and data processing techniques;
- Experimental control: The uncertainty of the analysis process is fundamentally in conflict with the controllability requirements of traditional experimental design;
- Evaluation granularity: Existing metrics (such as the System Usability Scale SUS [15]) can only evaluate overall usability and user acceptance, making it difficult to capture efficacy improvements at the analytical behavior level (such as discovery of reasoning paths, optimization of decision standards, user recognition of task completion and quality, etc.).

These challenges are not unique to VA systems but are methodological bottlenecks faced by all interactive systems containing various tools and means [16].

The phenomenon that VA research commonly adopts case study methods is likely due to the difficulties faced in implementing experimental methods. However, existing case study methods mostly focus on system functional descriptions, detailing how users combine various data processing and view tools to complete specific analysis tasks. Due to a lack of strict control over user tasks and tool usage, this case study method cannot establish causality between system design and analysis efficacy improvement, and the conclusions of individual studies on specific systems and tasks are also difficult to generalize into universal design principles. This limitation is essentially a concrete manifestation of the inherent flaws in validity and reliability of the case study method itself [17], [18] within the field of VA.

Interview methods essentially belong to the category of heuristic evaluation [19], which has specific value in evaluating specific dimensions of a system (such as interface usability, task execution flow). In VA research, interviews are often combined with case studies to obtain user subjective experience (e.g., tool learnability) and system improvement suggestions (e.g., functional flaws or missing features) through deep interviews with system users. Although this qualitative information enriches evaluation dimensions, interview methods share common methodological limitations with case studies: causal inference cannot be established due to the lack of effective control over variables, and descriptive data from local cases affects the reliability and scalability of research results.

In summary, the fundamental cognitive difference between simple interaction behavior and complex analytical behavior leads to systemic limitations of traditional HCI research methods when applied to visual analytics, a limitation that runs through several key research stages: requirement analysis, system design, and system efficacy evaluation. To break through this methodological predicament, it is necessary first to deeply parse the cognitive characteristics and process laws of analytical behavior, and then build a theoretical foundation targeting complex analytical behavior based on relevant theories to guide and develop effective requirement analysis, system design, and system evaluation.

## 4. Characteristics of Visual Analytics Behavior

We can understand the main characteristics of VA behavior starting from its existing definitions. Current academic concepts of VA are primarily based on two programmatic documents from the US and EU. The US definition originates from the book Illuminating the Path: The R&D Agenda for Visual Analytics [20], which defines VA as the science of analytical reasoning supported by interactive visual interfaces. But this definition is too simple: visual is interactive visualization, and analytics is analytical reasoning. The EU white paper Mastering the Information Age: Solving Problems with Visual Analytics [21] provides a more explicit definition: VA supports users' understanding of massive complex data and reasoning and decision-making on this basis by fusing automated data analysis and interactive visualization technologies.

Both definitions reveal that the core behaviors of VA include understanding, reasoning, and decision-making. In cognitive psychology, the cognitive processes involved in these behaviors are called complex cognition [22].

### 4.1 Complex Cognitive Behavior in Visual Analytics

In the field of complex cognition, activities such as learning, reasoning, problem-solving, and decision-making are recognized as typical complex cognitive activities [23], [24]. This recognition highly coincides with the EU white paper's definition of VA: the white paper title itself takes problem-solving as the ultimate goal of VA and explicitly constructs "understanding, reasoning, and decision-making" as the main cognitive means to achieve the goal.

Based on the dual perspectives of complex cognition theory and visual analytics, we can summarize user behaviors supported by VA systems into two main types: problem-solving and understanding. This classification both retains the theoretical framework of complex cognition research and highlights the research focus of the VA field.

We have temporarily not analyzed reasoning, learning, and decision-making as separate analytical behaviors here. This processing is based on the following considerations: first, in complex cognition theory, reasoning is considered an intermediate transitional behavior [25] that needs to depend on other cognitive behaviors such as learning, problem-solving, or decision-making, playing a role in supporting them. Second, although the VA field already has research supporting instructional behavior [3], it has not traditionally taken learning behavior itself as a core research object, possibly because typical learning behavior usually occurs in structured educational scenarios with clear knowledge acquisition goals, whereas "understanding" behavior involved in VA is often a sub-process of problem-solving or decision-making. It is worth noting that some recent research works on VA systems for explainable AI are essentially targeting learning behavior (i.e., acquiring new information and constructing new knowledge): for example, in a review by Yuan et al. on machine learning-oriented VA research [26], more than half (144) of 270 papers involve understanding of model results; however, such research rarely explicitly utilizes learning theory to guide user behavior analysis and system design.

### 4.1.1 Problem-Solving Behavior

The essence of cognitive behavior based on problem-solving is a cognitive process approaching a target result from initial conditions. If both the initial conditions and target result of a problem are clear and unambiguous, then the problem is a well-structured problem; otherwise, it is an ill-structured problem. Some well-structured problems can be solved with algorithms. But all ill-structured problems require human intervention to clarify vague and ambiguous conditions.

Problems targeted by VA systems usually are ill-structured problems containing many uncertainties. Taking a VA system for time-series data as an example [27], the user's goal is causal analysis of data, but finding the cause of a certain phenomenon is a very vague goal, and different hypotheses lead to different analysis strategies. The purpose of the VA system is to help users analyze and compare different hypotheses and strategies.

### 4.1.2 Decision-Making Behavior

As a typical complex cognitive behavior, two viewpoints exist in decision-making research: some literature believes decision-making and problem-solving have similarities in cognitive mechanisms [28], while other research emphasizes the essential differences between the two [29]. From the perspective of VA, we tend to view decision-making as a specific type of problem-solving behavior because decision-making behavior in VA systems usually manifests as generating candidate options based on given data and then filtering final options through some preset conditions. This process can view the final option as the target state of a problem, and option generation and evaluation as the problem-solving process.

Taking a football lineup VA system as an example [30]: the system can help coaches select a best lineup based on a specific opponent. During the decision process, the system can generate various lineups based on given information (such as situations of the opponent and own side), and then the coach selects the best lineup based on some criteria (such as players' states, degree of coordination, etc.). This process has initial conditions—various information about the opponent and own side—and a target result—the final lineup; selecting the best lineup is finding a solution based on given conditions. Of course, this decision problem is an ill-structured problem because the goal of "best lineup" is not very clear, as no criteria defining "best" are provided, requiring the coach to rely on personal experience to make a judgment.

Based on analysis of 230 VA papers, we found that VA system research mostly focuses on problem-solving type behaviors (148 papers), so we take problem-solving as the starting point for understanding VA behavior here.

### 4.2 Theories of Problem-Solving

Under the modern scientific paradigm, cognitive psychology has developed a mature theoretical system for problem-solving [31], providing a systematic theoretical framework and methodological support for in-depth research into the cognitive mechanisms of problem-solving.

### 4.2.1 Problem States, Problem Space, and Operators

A given problem usually contains two states: the initial state as the starting point (usually including given conditions, data) and the goal state as the purpose (i.e., the result). Problem-solving is constructing a series of logically coherent intermediate states based on a specific strategy to systematically connect the initial state and the goal state. These intermediate states, along with the initial and goal states, are collectively called problem states, and the collection of all problem states constitutes a complete problem state space. The process of problem-solving can be seen as a process of searching and selecting intermediate states within the problem space. Usually, multiple reasonable paths exist within a space, i.e., multiple problem-solving strategies. For example, when looking at a math proof from problem-solving theory, the given conditions of the problem are the initial state, the result to be proven is the goal state, and the intermediate steps of a reasonable proof are intermediate states. This problem may have multiple proof methods, and intermediate states involved in different methods may differ. The initial and goal states of the problem, along with all possible intermediate steps involved in proof methods, constitute the problem space for this problem. Proving this problem is searching for a path in the problem space that conforms to mathematical logic and norms to connect related problem states, and effective paths are often more than one.

Under a solution, when making a reasonable and logical transition from one intermediate state to another, it is usually necessary to rely on some means or tools, which are called operators in problem-solving theory. For example, proving a math problem needs to rely on some axioms and theorems, and these axioms and theorems are the operators for solving this problem.

Figure 1 shows the relationship between problem states, problem space, and operators. In this figure, two paths connecting the initial and goal states represent two problem-solving strategies.

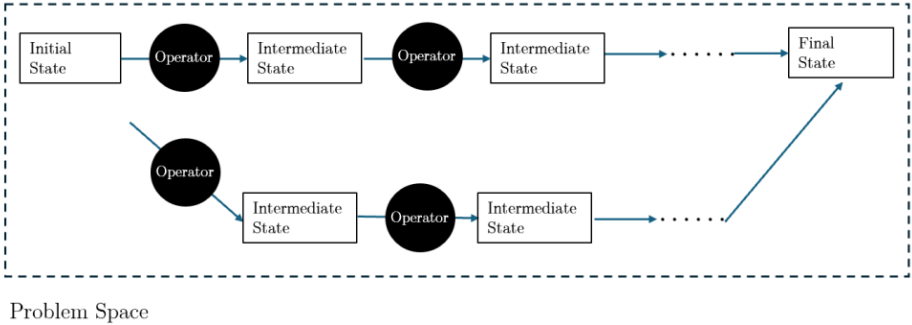

**Figure 1: Problem Space, Problem States, and Operators**

In most cases, problem-solving uses known operators; however, in some cases, existing operators cannot meet requirements, and it is necessary to create new operators. For example, if a math proof first needs to prove a new theorem and then complete the proof on the basis of this new theorem, then this new theorem is a new operator. But the new theorem needs to be strictly proven before it can be applied.

### 4.2.2 Problem Difficulty, Well-structured and Ill-structured Problems

The difficulty of a problem depends on various factors. One factor is the structuredness of the problem [32]. Well-structured problems have clear and reasonable initial and goal states, based on which the problem space can be clearly defined, and solutions can be obtained by determining intermediate states and related paths. Math problems in textbooks are basically well-structured problems because their initial conditions are basically sufficient and necessary, and goals are clear. Many well-structured problems have known solutions that can be executed automatically through tools, i.e., automation of the problem-solving process. For example, many math optimization problems and geometry problems can be solved or proven through programs.

Conversely, ill-structured problems have characteristics of problem space uncertainty such as lack of sufficient conditions, vague or even contradictory goals, creating difficulty in determining intermediate states and solution paths. In this case, obviously, automated problem-solving tools cannot be used.

Handling ill-structured problems usually requires intervention from an objective subject, making the problem well-structured by making some changes to the initial or goal states, or introducing local conditions and constraints based on specific situations when encountering obstacles while searching a path in an uncertain problem space. These handling means often depend not only on objective knowledge systems but are also influenced by subjective individual experience and values.

Ill-structured problems are usually regarded as more challenging due to their inherent uncertainty, but well-structured problems can also become difficult problems. There are many reasons leading a well-structured problem to become difficult: it could be that other intermediate states besides initial and goal states cannot be determined, leading to an inability to construct the problem space. When we face a math problem without knowing how to prove it, we fall into this predicament. Another situation is that even if a perfect space exists and some intermediate states are known, finding a reasonable path connecting these intermediate states faces difficulty. Sometimes the difficulty appears in the transition from one intermediate state to the next; for example, in math problem proof, the bottleneck of proof is the inability to establish logical cohesion between two important steps.

Whether solving easy or difficult problems, it involves searching the problem space, and methods for searching the problem space vary from person to person: experienced experts usually adopt a forward reasoning method from initial state to goal state, while beginners tend more toward a backward reasoning method from the goal [31].

### 4.2.3 The Cognitive Process of Problem-Solving

Cognitive processes involved in problem-solving usually include several key stages: problem identification and representation, strategy selection, execution, and evaluation. Problem-solving begins with identification and representation of the problem, i.e., the subjective individual needs to understand the structure and goal of the problem. According to Newell and Simon's "General Problem Solver" model [33], problem representation depends on the cognitive subject's understanding of important factors like the initial state, goal state, and possible paths. Wrong problem representation leads to invalid solution strategies, so this stage is crucial.

After clarifying the problem structure, the subjective individual selects an appropriate solution strategy. Common strategies include systematic, step-clear solutions and heuristic solutions based on experience. For example, the means-ends analysis method is a heuristic strategy whose goal is to reduce the difference between the current problem state and the final goal state. Additionally, analogical reasoning is also a common strategy, which draws on solutions to similar past problems to guide the current strategy.

The strategy execution stage involves complex cognitive processes, primarily relying on metacognition for monitoring functions. Metacognition can be defined as the cognitive subject's reflective cognition of its own cognitive process, i.e., awareness, monitoring, and regulation of complex cognitive processes such as problem-solving [34]. This metacognitive ability ultimately realizes the effective completion of cognitive tasks by ensuring the adaptability and flexibility of strategy implementation.

The evaluation and feedback stage focuses on assessing the effectiveness of the final solution. If results do not meet expectations, it is necessary to readjust problem representation or strategy, forming an iterative problem-solving process.

These problem-solving theories provide important theories and methods for VA system research. We rely on relevant theories and concepts of problem-solving (such as problem state space, operators, ill-structured problems, problem-solving strategies, etc.) to construct a cognitive behavior analysis framework applicable to VA.

## 5. VA Research Methods Based on Problem-Solving

Viewing VA behavior as a complex cognitive behavior of problem-solving can help us better understand the goals and tasks of various links in VA system research, thereby improving research schemes. Here, we take the effective support of problem-solving by VA systems as a starting point to discuss factors that need consideration in major research links. We focus on three aspects: the purpose of user requirement analysis, the utility of system design elements, and the emphasis of system evaluation.

### 5.1 Requirement Analysis: Focusing on the Cognitive Process of Problem-Solving

In the requirement analysis stage, regardless of the method used, researchers and designers must clarify what kind of user behavior needs to be observed and analyzed. Current common methods usually focus on aspects such as how users complete their work, what the existing interaction processes are when users work, and what problems they frequently encounter.

When the goal of system design is to help users solve problems, especially the difficulties of ill-structured problems, merely focusing on users' routine workflows and interaction methods is obviously insufficient. We need to gain a deep understanding of cognitive behavior and processes related to problem-solving when users solve a difficult problem. Although user interaction behavior logs (such as data used and menu command click history) can help us infer basic methods they adopt when solving a problem, these interaction record information are only external manifestations of user cognitive processes; some important steps of user internal cognitive processes (such as how users determine solving strategies, how they transform ill-structured problems, how they choose between different operators, etc.) are difficult to reflect in these data.

This requires us to consciously observe those behaviors related to problem-solving deeply when doing requirement analysis, understanding how they understand and interpret a problem (such as analysis of problem conditions and goals, understanding of problem space), how they handle ill-structured problems (such as addition and perfection of insufficient conditions, or clarification of vague goals), how they filter problem-solving strategies (such as exploration of possible solutions, screening of best solutions, recall of similar problems encountered in the past), etc.

Deep understanding of users' internal cognitive processes can adopt various research methods. If adopting the questionnaire survey method, deep analysis of problems targeted by the system and various obstacles likely encountered in solving these problems is needed in the questionnaire design stage to ensure the clarity, targeting, and effectiveness of questionnaire questions.

More open interview survey methods can also be adopted, giving users space to play freely and describe their problem-solving process specifically. This interview method requires the interviewer first to have a certain understanding of the user's work to correctly understand the effective means they adopt in the problem-solving process; second, the interviewer needs sharp insight to capture key information related to understanding and solving problems from user narratives and track or question further regarding this information to gain deep understanding of internal cognitive processes.

When adopting the field observation method, users need to use the think aloud method to provide real-time verbal descriptions of the thinking and cognitive processes guiding their interactive behavior. Through analysis of user verbal content, internal cognitive processes can be inferred. But when using this method, attention should be paid to deviations between user behavior and verbal descriptions as well as operational errors; when such problems appear, users should be asked to clarify or correct in a timely manner to ensure correct execution of problem-solving strategies.

In short, from the perspective of problem-solving, work in user requirements and analysis should focus on those behaviors and internal cognitive processes related to problem-solving, striving to deeply understand how users interpret a problem (such as cognition of problem space, determination of problem essence) and how they formulate solutions (such as selection of similar problems, exploration of possible alternative solutions, screening of the final solution). These information can reveal strategies, methods, and tools commonly used by users when facing such problems, as well as their cognitive resource allocation in related steps. On this basis, user requirements can be further summarized.

What needs emphasis here is that user requirements should be requirements at the problem-solving level involving problem-solving related factors (such as problem space, strategy, analogous problems, operators, etc.). For requirements targeting certain specific views or interaction tools, they should not only be limited to describing user expectations for them but should further elucidate their role and significance in problem-solving from the perspective of problem-solving. Design schemes proposed based on such user requirements will be more effective for improving the efficacy of complex analytical behavior because they target assisting users in improving the way they solve problems, rather than being limited only to certain specific operations in the process.

## 5.2 System Design from a Problem-Solving Perspective

Problem-solving theory can also help us re-recognize the utility of system design, gaining deep understanding of the roles of various system components and interaction tools in the problem-solving process.

### 5.2.1 Algorithms/Views: Operators in Problem-Solving

From the perspective of problem-solving, we can view both data processing algorithms and views as indispensable operators in the problem-solving process and accurately grasp the role and innovation of relevant design work.

Operators are very important in the problem-solving process, but solving a problem does not necessarily require new operators unless existing ones cannot meet requirements. In VA systems, data processing algorithms can be seen as a type of operator whose utility is transforming data from one state to another. For this link, if existing algorithms can achieve the goal (such as obtaining correct transformation results), then integrating the algorithm into the system is enough.

In some cases, when known algorithms cannot meet requirements, it is necessary to improve existing algorithms or even create new ones. From the perspective of problem-solving, a new algorithm is a new operator, and a new operator will have an impact on the related problem-solving process. Therefore, the importance of this work is no longer limited to the design innovation of a system component but lies more in its change to a problem-solving process (such as a new operator leading to a new solution, a new operator greatly shortening the problem-solving path, etc.); thus evaluation of this work should examine its impact on a problem-solving process from a deeper theoretical level, rather than being limited only to comparing efficiency with other similar algorithms.

For view design, we can similarly understand its role from the perspective of operators. If known views can help users transfer from one problem state to the next, then there is no need to design new views. Only when existing views cannot meet requirements is it necessary to improve existing views or design new ones. Similarly, such improvement and innovation will affect the problem-solving process.

The help of views for problem-solving is multi-faceted. First, views can directly display operator information (such as available algorithms and views) or state information (such as data) needed for problem-solving on the workspace, so users can conveniently find corresponding tools. The utility of views in this regard coincides with a consensus in the field of interaction design: externalizing certain interaction-related cognitive information (externalization) can reduce the memory burden in cognitive processes.

Another significance of views for problem-solving is that graphics can transform some abstract reasoning processes into comparisons and transformations of spatial structures (such as using histograms to simplify abstract numerical comparisons into comparisons of spatial object sizes). This transformation can also affect problem-solving strategies and efficiency.

### 5.2.2 Interaction: Supporting the Cognitive Process of Problem-Solving

When solving a problem, we need related declarative knowledge (such as problem states, operators, similar past problems, etc.) and procedural knowledge (such as usage methods of operators, transformation methods between different problem states, steps to solve similar problems, etc.), relying on short-term memory to process the current problem state and depending on long-term memory to obtain relevant similar past problems and solutions. How to establish the link between short-term memory and long-term memory can be achieved by visualizing relevant cues, which requires visualization design to consider the link between the current problem state and possible similar problems, and this link is based on user interaction behavior, so it is dynamic. Therefore, system design should strive to help users smoothly explore the problem space, reaching the goal state from the initial state.

Most current interaction designs stay at the level of supporting interaction between users and operators at the operational level, while lacking support for macro-level problem-solving strategy exploration behavior. For example, when facing an ill-structured problem, users need to understand various elements of the problem space to analyze where the crux of the ill-structuredness lies. When facing a difficult well-structured problem, users may need to understand numerous problem states in the problem space to find the next problem state and possible connection paths. In both cases, users perhaps expect tools to assist them in various explorations of the problem space, such as viewing changes in problem space states after adding different initial conditions, testing the impact of different operators on the links between problem states, etc. Current VA systems rarely exhibit work in this regard.

The problem-solving process frequently uses the analogy method, borrowing strategies from solving similar past problems to solve the current problem. But recalling and searching for similar problems based on the current problem state is a cognitive difficulty because many similar problems, although essentially the same, have different surface characteristics; users need certain knowledge and experience to see through surface phenomena to the essence of the problem [31]. Current VA systems also lack effective research and tools in how to help users clarify the essence of a problem.

Additionally, strategies, processes, and results of problem-solving have strong correlations with user knowledge, skills, goals, etc. Analysis tools expected by different users may differ greatly, and their explorations of problem space are also not identical. For example, existing cognitive theory shows that experienced analysts are usually familiar with relevant problem concepts and states, so they focus on relying on procedural knowledge to solve problems, while beginners need to first familiarize themselves with relevant concepts and then rely on relationships between concepts to find solutions. Therefore, systems should consider how to support personalized problem-solving methods of different users at the interaction level, rather than providing a set of standardized tools or work templates for users to adapt to.

Obstacles to problem-solving are numerous. For an ill-structured problem, the greatest challenge is how to make it well-structured. Since everyone's knowledge, experience, goals, etc., are different, they have different strategies for how to make the same problem well-structured; therefore interaction design should consider helping users smoothly explore a problem space, understanding various elements of the problem space, thereby helping them understand where the crux of ill-structuredness lies, and then decide what measures to take for making it well-structured (such as enriching initial conditions, clarifying the final goal, etc.).

For a well-structured but problem space-complex or path-long problem, interaction design, besides supporting user exploration of problem space, should also consider assisting users in managing various problems encountered during the path exploration process, such as saving important intermediate states and displaying relationships between related states in data and tasks.

### 5.3 System Evaluation: Measurement of Complex Cognition

Based on problem-solving theory, we can also upgrade VA system evaluation means, considering the impact of the system on complex cognitive processes on top of traditional tool usability.

When using the case study method, we can focus on understanding how users use the system to explore the problem space. For example, we can analyze how the system helps users manage numerous problem states, how it assists users in exploring and comparing different problem-solving paths, how it helps users search for similar problems, etc.

User interviews should also focus on questions related to problem-solving strategies and methods. For example, interview questions can involve which step in the analysis (i.e., transition from one problem state to another) was key, what algorithm and view tools were relied upon, what methods were adopted to define the problem more explicitly (i.e., making an ill-structured problem well-structured), how tools provided by the system assisted these links, etc.

Besides gathering problem-solving related information in common case studies and interviews, we can also obtain relevant information through questionnaires. Questions in this regard should focus on complex cognitive measurements at the problem-solving level, such as user confidence in the solution obtained by relying on the system, and comparing the user's level of understanding and familiarity with the problem before and after using the system.

Complex cognition theory also provides new measurement means for experimental verification methods. For example, Ackerman et al. [35] proposed that in experimental methods testing the effectiveness of interactive systems, measurements such as user resolution (the degree of discrimination between correct and incorrect solutions) and confidence in problem solutions can be added to examine the system's impact on user metacognitive ability.

Compared with traditional evaluation methods based on usability metrics and limited to completion of certain specific tasks, these evaluation means focused on general problem-solving processes and metacognitive characteristics can improve the external validity of research results and enhance the contribution of research at the theoretical level.

In summary, examining VA research from the perspective of problem-solving can provide a theoretical framework with more cognitive depth for requirement analysis, system design, and evaluation: requirement analysis can break through observation of user surface interaction behavior and delve into internal cognitive processes (such as problem space construction, transformation of ill-structured problems, strategy screening logic, etc.); system design can further clarify the support of each component in the system for the problem-solving process (such as data processing algorithms and views as key operators to inspire or assist transformation of problem states and optimization of solving paths), as well as the role of interaction design in supporting dynamic problem space exploration and adaptation of related strategies; system evaluation helps break through traditional usability measurements and build an indicator system oriented toward complex cognitive processes (such as the system's support capacity for key steps like problem space exploration and strategy screening, and metrics of user satisfaction and trust in results).

## 6. Application of Other Complex Cognition Theories

Our discussion in the previous section focused on related theories of problem-solving because critical problems focused on by most VA system research can be summarized as problem-solving type challenges. But we believe our method can be applied to analyze the relationship between other complex cognitive behavior theories and VA behavior. Here we outline the guiding significance of related theories of decision-making and learning for VA system research.

Decision-making behavior is also a branch deeply studied in the field of cognitive psychology [36]. Similar to problem-solving behavior, decision-making behavior is also a multi-stage process, including stages of problem identification and goal determination, information search and option generation, option evaluation and trade-off, decision execution and evaluation, as well as result feedback and learning, and its cognitive process also involves metacognition. Through deep understanding of decision-making theory, we can also establish correspondence between decision-making theory and analysis behavior, further guiding research on VA systems oriented toward decision-making behavior. We can guide requirement analysis according to the cognitive characteristics of decision-making, learning what challenges users face in the decision process related to elements of decision behavior (such as lack of clear decision purpose, lack of necessary decision options and evaluation criteria, etc.), and then assist key stages of decision-making through system design (such as helping generate required options and displaying key evaluation criteria through algorithms and views). In evaluation, we can still adopt measurements at the metacognitive level to evaluate the efficacy of the system on general decision processes and metacognitive capacity.

Additionally, we also need to focus on unique characteristics of decision behavior, exploring the impact of VA tools on these characteristics. For example, decision theory points out that decision results will be influenced by many cognitive biases of the decision-maker [37], and Wall [38] et al. also analyzed some major cognitive biases in the VA process. Based on this understanding, system design, while supporting decision elements and processes, also needs to consider how to reduce or avoid the impact of cognitive biases on decision-making through technical means (such as designing algorithms to capture cognitive biases in the analysis process, using views to display decision errors likely generated by cognitive biases, etc.), allowing users to detect potential cognitive biases, correct them in a timely manner, and make correct decisions.

Metacognition theory also provides a new perspective for research directions such as VA systems targeting learning and storytelling VA systems. Theories related to learning, such as self-regulated learning theory [39], can help researchers and designers deeply understand user learning processes and behavioral characteristics, and accordingly design appropriate tools to enhance learning efficacy. For example, under the guidance of these theoretical frameworks, research on VA systems for explainable AI can focus on understanding user goals, methods, and obstacles in the process of learning AI models, designing views and interaction means based on learning theory to help users effectively master key concepts, indicators, workflows, etc., in the model, and then verifying system effectiveness by measuring changes in user knowledge structures before and after using the system [40]. Similarly, in the design of storytelling VA systems [41], theories related to creative behavior [42] can be borrowed to guide design and evaluation.

## 7. Conclusion

By analyzing mainstream methods in current VA system research, we pointed out that these methods based on simple cognitive behavior have a mismatch with research goals oriented toward complex analytical behavior. Taking problem-solving theory in the field of complex cognition as an example, we elucidated how to establish the relationship between complex cognition theory and VA behavior, and the guiding significance of these theories for major means of VA research.

In short, we believe the deep fusion of complex cognition theory and VA practice can help us explore a new paradigm for VA system research. By deeply understanding theories related to complex cognition in fields like problem-solving, decision-making, and learning, we can construct a methodological system for VA system design with higher internal and external validity.